\documentclass[prl,
twocolumn,
showpacs,showkeys,floatfix]{revtex4}
\usepackage{latexsym,amssymb,amsmath,amsfonts}
\usepackage{graphicx}
\usepackage{bm}
\newcommand{\beq}{\begin{equation}}
\newcommand{\eeq}{\end{equation}}
\newcommand{\bc}{\begin{center}}
\newcommand{\ec}{\end{center}}
\newcommand{\eeqa}{\end{eqnarray}}
\newcommand{\beqa}{\begin{eqnarray}}
\newcommand{\no}{\noindent}

\newcommand{\ra}{\rightarrow}

\newcommand{\al}{\alpha}
\newcommand{\be}{\beta}

\newcommand{\ph}{\phi}

\newcommand{\ed}{\end{document} }

\begin{document}

\raggedright
\parindent3em

\title{Dark Entropy}
\author{Richard T. Hammond}
\email{rhammond@email.unc.edu}
\affiliation{Department of Physics,
University of North Carolina at Chapel Hill,
and the
Army Research Office
Research Triangle Park, North Carolina, 27703}
\author{Terry~Pilling}
\email{Terry.Pilling@ndsu.edu}
\affiliation{Department of Physics,
North Dakota State University,
Fargo, North Dakota 58105}

\date{\today}


\begin{abstract}

We examine the consequences of a universe with a non-constant
cosmological term in Einstein's equations and find that the
Bianchi identities reduce to the first law of thermodynamics
when cosmological term is identified as being proportional to 
the entropy density of the universe. This means that gravitating
dark energy can be viewed as entropy, but more, the holographic
principle along with the known expansion of the universe 
indicates that the entropy of the universe is growing with
time and this leads to a cosmic repulsion that also grows with time. 
Direct implications of this result are calculated and shown to 
be in good accord with recent observational data.

\end{abstract}

\maketitle

One of the biggest mysteries in present 
day cosmology is the origin and identity of the so-called 
``dark energy'' of the universe which is necessary to account 
for the accelerated expansion \cite{knop,reiss,virey}. 
The most promising candidate for this dark energy is a cosmological 
constant, although any attempts to calculate the observed value 
using quantum theory results in a value that is many orders of 
magnitude larger than it should be. 
It therefore seems worthwhile to check whether the cosmological 
constant could arise from some other source rather than the vacuum 
energy. One possibility comes from considering a non-constant
cosmological term, $\Lambda(t)$ \cite{varyingL}. This
term would appear in Einstein's equation in exactly the same
way as the usual cosmological constant except that it varies
with time. It is well known however that the Bianchi identity
along with the fact that the energy-momentum tensor is covariantly
conserved, imply that $\Lambda = constant$. However, this
conservation law merely regulates the exchange of energy
and momentum between source fields and would fail to hold
in the case that there is an independent source or sink of
matter or energy in the universe. In this case these conservation
equations simply become equations regulating the exchange of 
energy and momentum between matter, gravitational energy and the 
cosmological term, rather than just matter and gravitational
energy alone. 
It was shown by many authors \cite{varyingL} that an increasing 
cosmological term can account for
the accelerated expansion of the universe but, as with all
other solutions involving a cosmological constant, no physical
identity is postulated for this dark energy, no mechanism 
provided for its variability with time and therefore there 
is no way for these models to {\it predict}, based on known
physical principles obeyed by $\Lambda(t)$, the precise details 
of the accelerated expansion that is being observed. 
Here, on the other hand, we do exactly that. We find the 
physical identity of the time-varying cosmological term. In
particular, we show that dark energy as described by a 
non-constant cosmological term and leading to a shift in
the energy density and pressure of the universe can be 
identified with entropy. We use this identification
to predict the details of the current expansion.

We have examined Einstein's equations in the presence of a
non-constant cosmological term and found that the Bianchi identity 
reduces to a first law of thermodynamics if one identifies 
the cosmological term as proportional to the entropy density 
of the universe. This can be combined with the holographic 
principle to make specific predictions as to
the current evolution of the universe.

To see how the holographic principle enters into the picture,
let us first recall the example of a black hole.
A black hole is the smallest possible object of a given mass. 
It is also the object with the highest entropy in a given volume 
of space. If  matter is added to a black hole its entropy increases and its
size, the area of its event horizon, increases proportionally
\cite{bekenstein,hawking}.
Thus, after gravitational collapse, the size of the object is 
defined as the horizon size, or entropy, rather than the extent 
of the matter content. 
The idea that the maximum entropy of a region of space is 
proportional to the surface area surrounding that region has
been generalized beyond black holes to the rest of universe
in the so-called holographic principle \cite{thooft1993,susskind1995}.
This principle was used to resolve the so-called black hole
information paradox and is generalized to the entire universe.
Let us assume, as in the holographic principle, that the 
entropy of a volume of space is proportional to the area of the
boundary of that region in terms of the number of Planck sized 
cells on that boundary.

The universe began a finite time in the past with a big bang.
Since then the fabric of spacetime has been expanding.  
On the other hand, bound states
such as elementary particles and galaxies are not expanding, instead
they are moving apart as the fabric of spacetime that contains
them expands. The constants of nature $G$, $\hbar$ and $c$ 
are presumably unchanging, so that the fundamental length scale 
which is formed from them, the Planck length, is also
unchanging. The Planck length is therefore tied to the scale of
the bound states -- the elementary particles -- rather than the
fabric of the spacetime itself. But the entropy-area relationship
built into the holographic principle and also black holes tells us 
that the entropy of a region is proportional to the area of its
boundary, i.e. {\it the number of Planck-sized cells}. 
This implies that as the universe expanded in the time since the
big bang, the number of Planck sized cells on its boundary must 
increase since the area is increasing but the fundamental size of 
the cells is not. Thus, the entropy of the universe must increase 
with the expansion and this entropy is `dark' in that it comes 
from the changing boundary area rather than the matter content 
of the universe. 

As long as the energy content dominates the dark entropy content, the 
expansion will slow as the matter tries to pull the universe back 
in on itself. Eventually however, the increasing push from the entropy 
will exceed the decreasing pull from the energy and the expansion of 
the universe will begin to accelerate.
This accelerated expansion is exactly what astronomers are currently
seeing.

In the following section we  examine Einstein's 
equation in a Friedmann-Robertson-Walker (FRW) background with the addition 
of a non-constant cosmological term \cite{hammond99} and show 
how dark energy arises naturally. We then show how this
dark energy is identified with entropy where the entropy is
proportional to the boundary area of the universe in units
of the Planck area. 
We then explore the observational effects of this dark entropy 
and show that it solves the problem of the anomalous acceleration
in the universe, yielding a deceleration parameter that was originally 
positive but has a current negative value in approximate agreement with 
recent observations. Natural units are used until the end.

The above arguments imply that the Einstein-Hilbert action must be 
generalized: the simplest possibility being

\begin{equation}\label{action}
I=\int d^4x\sqrt{-g}\left(\frac{R}{16\pi} + \Lambda \right)+I_m
\end{equation}
where  $\Lambda$ is proportional to the entropy density and $I_m$ 
describes matter in the usual way.

Such an idea is not without foundation. 't Hooft has argued from the 
holographic principle, emphasizing that the information in a volume is 
proportional to the surface area \cite{thooft1993}. Moreover it is well known 
that entropy enlists surface terms to the action, but here we assume the 
more direct idea that it is the entropy itself that is included. 
Also, since it is believed that the deceleration was once positive, but 
became negative some time in the past, we are looking for a field that 
grows with time (in magnitude). In an expanding universe most cosmological 
fields decrease with time, but entropy always increases. 

Taking all these clues as hints toward (\ref{action}), we have, assuming 
the entropy density is a scalar quantity independent of the metric tensor,

\beq
G^{\mu\nu}= 8 \pi T^{\mu\nu} + 8 \pi \Lambda g^{\mu\nu}
.\eeq
In order to comply with the holographic principle, it is assumed that 
the entropy of space is proportional to the area, i.e., 
\beq\label{S}
S = \gamma A/4L_P^2,
\eeq
where $\gamma $ is an unknown proportionality constant (which reduces
to the Boltzmann constant in the case of a black hole).

For cosmology, let us consider the Robertson Walker metric and take 
the energy momentum tensor to be that of a perfect fluid:
\beq
T^{\mu\nu} = (\rho + p) v^\mu v^\nu - p g^{\mu\nu}.
\eeq

\no The 0-0 field equation becomes
\beq\label{00}
\frac{3}{a^2}(\dot{a}^2 + 1) = 8 \pi (\rho + \Lambda)
\eeq
and the rest are equivalent to,
\beq\label{11}
\frac{1}{a^2} (\dot{a}^2 + 1 + 2a\ddot{a}) = -8 \pi (p - \Lambda)
.\eeq
In fact, as is well known, (\ref{00}) is equivalent to  (\ref{11}) provided the
Bianchi identity holds, which gives
\beq\label{bi}
T^{\mu\nu}_{\ \ ;\nu} + \Lambda^{,\mu} = 0.
\eeq

Using the line element of a Friedman-Robertson-Walker cosmology
\begin{equation}
\label{FRW}
ds^2 = dt^2 - a^2\left( d\chi^2 +\sin^2\chi(d\theta^2+\sin^2\theta d\ph^2)  \right) 
\end{equation}

\no where $V$ is the volume of the universe ($2\pi^2a^3$).
Defining $M=\rho V$ we can write (\ref{bi})
instead as
\beq\label{bi2}
dM = -V d\Lambda - p dV
\eeq
Comparing this with the first law of thermodynamics we see a strong 
motive for the association of entropy with the cosmological term. In the 
relativistic form of the first law, $dU$ is replaced with $dM$, and in 
fact (\ref{bi2}) is identical to the first law if we take $-Vd\Lambda=TdS$. 
This works explicitly if $\Lambda$ is proportion to the entropy density, 
$\Lambda=KS/V$. With this (\ref{bi2}) becomes
\beq\label{2law}
dM = TdS - pdV
\eeq
with $T=K/2$.

This differs in application to the conventional first law in that
this applies to space itself. In fact, the above association yields a 
temperature of space $T\approx 4 \times 10^{-50}$K (which has nothing to do 
with the background radiation temperature).

Let us proceed for the case that the pressure $p$ is 
negligible: (\ref{2law}) can be integrated to give,
\beq
M = \frac{{\cal K} a^2}{16 L_P^2} + m
\eeq
where $m$ is a constant of integration and ${\cal K} \equiv 8 \pi \gamma K$.
Using the definition of $M$ we get
\beq
\rho = \frac{{\cal K}}{ 32 \pi^2  L_P^2 a} + \frac{m}{2 \pi^2 a^3}
.\eeq
It may be noted that by setting ${\cal K} = 0$ ($K=0$) the entropy terms 
are made to vanish. In this limiting case, everything reduces to the 
standard Friedman cosmology.

Now we may consider the cosmological implications. 
Units are non-dimensionalized using today's value of the Hubble 
constant $H_0$, i.e., $a \ra a H_0/c$ and $t \ra tH_0$ so we have
\beq
\dot{a}^2 + 1 = \frac{\alpha}{a} + \beta a
\eeq
and
\beq\label{q}
2 q H^2 = \frac{\alpha}{a^3} - \frac{\beta}{a}
\eeq
where, adopting cgs units for the moment, the dimensionless constants 
are $\al = 4mGH_0/3\pi c^3$, $\be =  {\cal K}c / 4\pi H_0 L_P^2$, 
and where $H = \dot{a}/a$ and $q$ is the deceleration parameter which 
can be computed from its definition, $q= -a\ddot{a}/\dot{a}^2$, or 
from (\ref{q}), which are equivalent.
Thus, there are two unknown constants,  $\al$ (from $m$) and 
${\be}$ (from $\gamma K$).
These constants may be found by comparing to the known constants of the 
Hubble constant today and the deceleration parameter. 

We can also define the equation of state parameter $w$ as follows. 
If
\beq
8\pi (\rho +\Lambda ) \equiv 8 \pi \rho_d
\eeq
and 
\beq
- 8\pi (p - \Lambda ) \equiv - 8 \pi p_d
,\eeq
we may define
\beq
\label{w}
w = \frac{p_d}{\rho_d}
,\eeq
which gives
\beq
qH^2 = \frac{4\pi}{3}(1+3w)\rho_d.
\eeq
This shows the known result that acceleration will occur for $3w<-1$. 
From (\ref{w}) we may also obtain
\beq
w = - \frac{1}{1+ \frac{\al}{2\beta} \left(\frac{\beta}{a} 
+ \frac{3\alpha}{a^3}\right)}.
\eeq

In Fig. 1 we plot the Hubble value and $q$ and $w$ along with the 
increasing radius parameter $a$.
The values of $\beta=2$ and $\alpha=1/4$ were chosen so that $H$ and $q$ 
are in line with recent observations (see refs. \cite{knop,reiss,virey}). 
For example, if $t_0=14$ billion years, then Fig. 1 gives 
$H(t_0)= 77$ km s$^{-1}$ mps$^{-1}$.
This also shows that $q$ became negative at $t\sim .35$, when the universe 
was about 5 billion years old and today enjoys the value of $q=-.55$. 
The graph also shows today's value of $w$, which gives $w= -0.85$.

In summary, the notion that a trapped surface contains entropy 
proportional to the area $S= \gamma A/4L_P^2$ is generalized to the 
assumption that it is a fundamental 
property of space. In fact, it is found that the Bianchi identities gives 
rise to a first law of thermodynamics if the cosmological constant is taken 
to be proportional to the entropy density. The equations of motion follow 
from the Bianchi identities even when other fields are present \cite{hammond2}, 
but in this case we have restricted the analysis to galaxies that are 
assumed to be at rest in the comoving coordinates. 
When applied to the cosmos, the theory contains two unknown constants, 
the coupling constant for the entropy density, and a constant that would 
reduce to the ``mass of the universe'' in the limit that the entropy vanishes. 
These are fixed by the Hubble constant and the deceleration parameter. 
Although only a closed space is considered here, an open and flat universe 
work as well, without changing the values of $\alpha$ or $\beta$ much. 
It is shown that the deceleration parameter is initially greater than 
zero but must become negative, with a current value of roughly $q=-0.55$, 
in agreement with current observations.  

\begin{widetext}
\begin{center}
\begin{figure}[ht]
\includegraphics{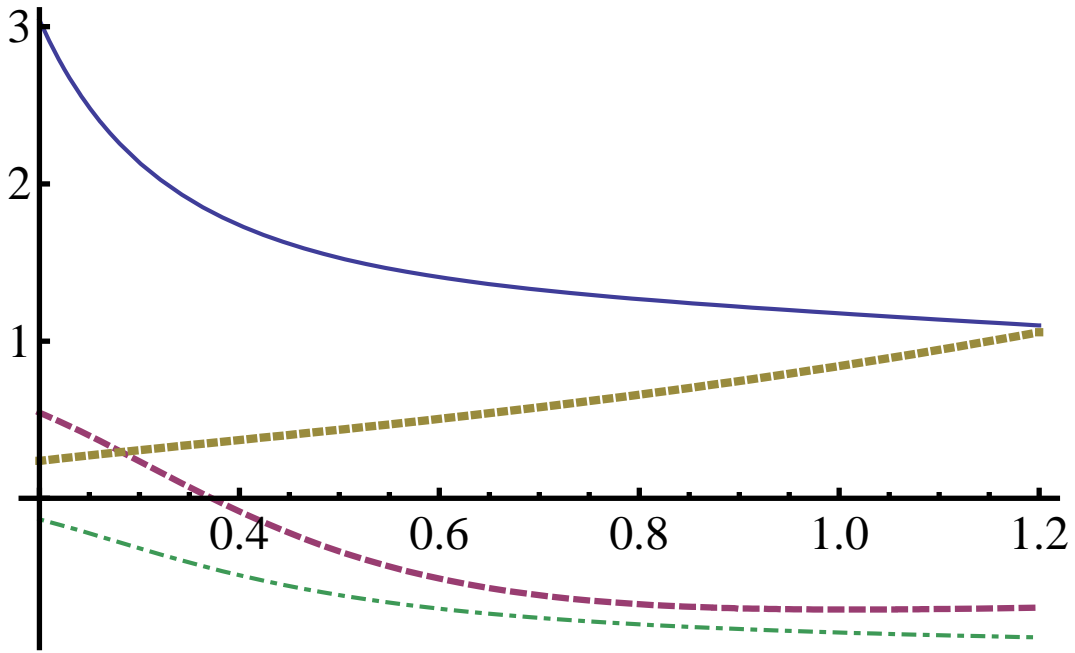}
\caption{Hubble's constant (solid), 
the increasing radius (dotted) of the universe,
the deceleration parameter (dashed), 
and $w$ (dot-dash) as a function of time for $\al=1/4$ and $\beta=2$.}
\end{figure} 
\end{center}
\end{widetext}

\end{document}